\documentclass{aastex62}
\usepackage{listings}
\usepackage{color}
\usepackage{float}
\usepackage{xspace}
\usepackage{inconsolata}

\definecolor{dkgreen}{rgb}{0,0.6,0}
\definecolor{gray}{rgb}{0.5,0.5,0.5}
\definecolor{mauve}{rgb}{0.58,0,0.82}

\graphicspath{{./}{figures/}}

\submitjournal{ApJ}

\shorttitle{TOAST, TEA, and More}
\shortauthors{McGlynn  et al.}

\begin{document}

\title{Octahedron-Based Projections as Intermediate Representations for Computer Imaging: TOAST, TEA and More.}

\correspondingauthor{Thomas McGlynn}
\email{Thomas.A.McGlynn@NASA.gov}

\author[0000-0003-3973-432X]{Thomas McGlynn}
\affil{NASA/Goddard Space Flight Center \\
Greenbelt, MD 20771, USA}

\author[0000-0003-2500-8984]{Jonathan Fay}
\affiliation{Microsoft Research \\
One Microsoft Way \\
Redmond, WA 98052, USA}

\author{Curtis Wong}
\affiliation{Microsoft Research \\
One Microsoft Way \\
Redmond, WA 98052, USA}

\author[0000-0001-9306-6049]{Philip Rosenfield}
\affiliation{American Astronomical Society\\
1667 K St NW Suite 800\\
Washington, DC 20006, USA}

\begin{abstract}

This paper defines and discusses a set of rectangular all-sky projections which have no singular points, notably the Tesselated Octahedral Adaptive Spherical Transformation (or TOAST) developed initially for the WorldWide Telescope (WWT).  These have proven to be useful as intermediate representations for imaging data where the application transforms dynamically from a standardized internal format to a specific format (projection, scaling, orientation, etc.) requested by the user.  TOAST is strongly related to the Hierarchical Triangular Mesh (HTM) pixelization and is particularly well adapted to the situations where one wishes to traverse a hierarchy of increasing resolution images.  Since it can be recursively computed using a very simple algorithm it is particularly adaptable to use by graphical processing units.

\end{abstract}


\section{Introduction} \label{sec:intro}
Hundreds of map projections have been developed over the course of many centuries \citep[e.g., see][]{Snyder1993} trying to represent a spherical surface despite the physical realities of publishing information on flat sheets of paper and monitors.  Different projections are designed to meet different goals:  the Mercator projection is designed to ease navigation, it accurately represents the directions between nearby points so that a sailor can steer a boat properly.  For this purpose, the projection's gross distortion of polar regions is a minor nit.   The Molleweide and Aitoff projections give pleasing all sky images with limited distortion.  The tangent plane or gnomonic projection is frequently used in astronomy since it often approximates the behavior of real small scale astronomical images.  Similarly the orthographic or sine projection arises naturally in images derived from interferometry.  The Hierarchical Equal Area Pixelization's (HEALPix, \citet{Gorski2005}) equal-area, and iso-latitude characteristics make it very attractive for the computation of the spherical harmonics used in the analysis of the cosmic microwave background.
% * <tom.mcglynn@nasa.gov> 2018-04-27T19:37:50.560Z:
%
% ^.

Computers have made it much easier to both define and use projections.  The widely used WCSLIB\footnote{http://www.atnf.csiro.au/people/mcalabre/WCS/wcslib/index.html} supports dozens of different projections.  With modern computers, users can rapidly project and transform images from one projection and coordinate system to another.    Computers can also provide access to very large images.  Systems can cache images at multiple resolutions, with a low resolution image of a large area (perhaps the entire sky) split into multiple tiles of higher resolution.  These tiles may themselves be split into higher resolution sub-tiles, forming a hierarchy of ever finer resolution.

The ability of software to rapidly transform data into any given output projection and coordinate system has established a new driver for projections: providing a flexible internal representation of the data that can be easily transformed into users' desired outputs.   It is possible to build systems which can transform directly from and to each of a set of known projections.  However, the number of transformations required goes up as the square of the number of projections involved.  A more practical approach can be to consider one fiducial representation and transform this to and from the other representations.  The intermediate frame is not intended to be `seen' by users.  The form of this intermediate projection would have different goals from those that have driven the development of earlier sky projections.  

This paper discusses a class of projections that have some particularly desirable characteristics for this purpose.  The remainder of this introduction discusses the characteristics we would like for an ideal intermediate projection.  In the next section we describe a set of projections based on transformation from the sphere to cubes. These were derived in a fashion similar to the projections we introduce below and can  help us understand how our new projections relate to older approaches. 

Section 3 shows a `topological' framework for how we build our class of projections, which use octahedrons rather than cubes.  This addresses how we cut the sphere to produce our image in the projection plane.   

The fourth section discusses three specific realizations of this topology including the Tessellated Octahedral Adaptive Spherical Transformation (TOAST).    Section 5 discusses specific aspects of the TOAST projection and why it is especially well adapted for use in GPU-based computations. Our conclusions note some of the areas where these projections are being used and are available and briefly explores the possibility of projections using other regular solids.

Consider images which we have stored in some standard intermediate projection where the data is going to be resampled for display into some other projection, scale, or coordinate system to meets a user's particular goals.  What makes a good intermediate projection?  We suggest four basic criteria.

\begin{enumerate}
\item The projection should be able to represent the entire sky in a single image.  If multiple images are required to represent the sky then software needs to worry about the cross-over points and which image to use for which point in the sky.  The projection should be easily adaptable to all sizes of image from the entire sky to a tiny area around a point source. 
\item The all-sky projection should be representable as a finite square or rectangle.  Storage can be allocated efficiently and much of our software assumes this organization.  If there are curvilinear boundaries then pixels along the border may be very difficult to deal with properly:  part of the pixel in the projected region and part outside. If we wish to tile it is trivial to split a rectangle into sub-tiles so a hierarchy of tiles at various resolutions can be easily supported.   
\item There should be no discontinuities or singularities in the projection.  If there is a point at which the transformation between the sphere and the projection is singular it is going to be much more difficult to accurately transform the region around the singularity.  One way to find discontinuities is looking at the Tissot Indicatrices for projections.  These show how small circles on the sphere are rendered in the projection plane.  If the area of the projected circles goes to 0 or infinity we have one class of singularity.  We can also have a singularity when the circles become infinitely long and thin, even though they may preserve area. 
\item The projection should be continuous across all boundaries – including the outer boundaries of an all sky image.  E.g., consider a standard Cartesian projection: While an all sky image projects to a finite rectangle, one can trivially extend the east and west boundaries by repeating the data cyclically.  Even if we happen to be near the eastern or western edges of a Cartesian map we can be assured that there is no real discontinuity at that edge.
\end{enumerate}
These four simple criteria rule out the projections in common use.   Table \ref{table:projections} lists the projections in Table 13 from \citet{Calabretta2002} and notes where they meet these criteria.

The orthographic/sine and gnomonic/tangent plane projections can only map half the sky. The Mercator projection cannot include the poles.   Most of the pleasing all-sky projections, Aitoff, Molleweide and the like are not rectangular.  The Plate-Caree or Cartesian projection meets that criterion but is singular at the poles as are the other cylindrical projections.   A stereographic image does still a little better, it is singular at only one pole, but it is not finite.    The quadralaterilized cube projections nearly meet all of the criteria but map the sky to multiple rectangles, not a single one.  None of the commonly used (and this table includes some uncommon ones) sky projections meet our criteria.

Due to its popularity we have added one projection to  the list in the \citet{Calabretta2002}, the Hierarchical Equal-Area Pixelization\footnote{http://healpix.jpl.nasa.gov} (HEALPix) suggested by \citet{Gorski2005}, \citep{Calabretta2007}).  The HEALPix representation can represent the entire sky and has no singularities, but the standard rendering is not rectangular.  The most compact representation is as a rectangle with serrated edges.
% * <tom.mcglynn@nasa.gov> 2018-04-27T19:40:52.079Z:
%
% ^.

\begin{deluxetable}{lcccc}\label{table:projections}
\tablecaption{Commonly Used Astronomical Projections}
\tablehead{
\colhead{Projection}	&
\colhead{Finite}	&	
\colhead{Rectangular}	&	
\colhead{Singularity Free}	&	
\colhead{Continuable}	\\
}
\startdata
Zenithal Perspective (AZP)	&	Yes	&	No	&	No	&	No	\\
Slant Zenithal Perspective (SZP)	&	Yes	&	No	&	No	&	No	\\
Gnomonic or Tangent (TAN)	&	No	&	No	&	No	&	No	\\
Stereographic (STG)	&	No	&	No	&	No	&	No	\\
Slant Orthographic or Sine (SIN)	&	No	&	No	&	No	&	No	\\
Zenithal Equidistant (ARC)	&	Yes	&	No	&	No	&	Yes	\\
Zenithal Polynomial (ZPN)	&	Varies	&	No	&	No	&	Yes	\\
Airy (AIR)	&	No	&	No	&	Yes	&	Yes	\\
Cylindrical perspective (CYP)	&	Yes	&	Yes	&	No	&	Yes	\\
Cylindrical equal area (CEA)	&	Yes	&	Yes	&	No	&	Yes	\\
Plate carree or Cartesian (CAR)	&	Yes	&	Yes	&	No	&	Yes	\\
Mercator (MER)	&	No	&	No	&	No	&	No	\\
Molleweide (MOL)	&	Yes	&	No	&	Yes	&	Yes	\\
Hammer Aitoff (AIT)	&	Yes	&	No	&	Yes	&	Yes	\\
Conic Perspective (COP)	&	Yes	&	No	&	Yes	&	No	\\
Conic Equal Area (COE)	&	Yes	&	No	&	Yes	&	No	\\
Conic Orthomorphic (COO)	&	Yes	&	No	&	Yes	&	No	\\
Bonne's Equal Area (BON)	&	Yes	&	No	&	Yes	&	No	\\
Polyconic (PCO)	&	Yes	&	No	&	Yes	&	No	\\
Tangential Spherical Cube (TSC)	&	Yes	&	No	&	Yes	&	No	\\
COBE Quadrilateralized Spherical Cube (CSC)	&	Yes	&	No	&	Yes	&	No	\\
Quadrilateralized Spherical Cube (QSC)	&	Yes	&	No	&	Yes	&	No	\\
HEALPix	&	Yes	&	No	&	Yes	&	No	\\
\enddata
\end{deluxetable}

\section{Lessons of cube-based projections}
We noted that one set of projections that come close to meeting the requirements are projections where the sphere is projected onto the six facets of an enclosing cube.   This kind of projection was popularized with COBE Spherical Cube (CSC) projection.  Several variants have been developed and three are shown in Table \ref{table:projections}.   The projections we discuss below have many similarities to the cubic projections, so we will discuss these in a bit more detail.

If we look at the cube projections as a class we can view them as having two elements, a common topological element where we determine which face to map a given element of the sky to, and a detailed transformation rule within the face which differs among the projections.  

\begin{figure}
\plotone{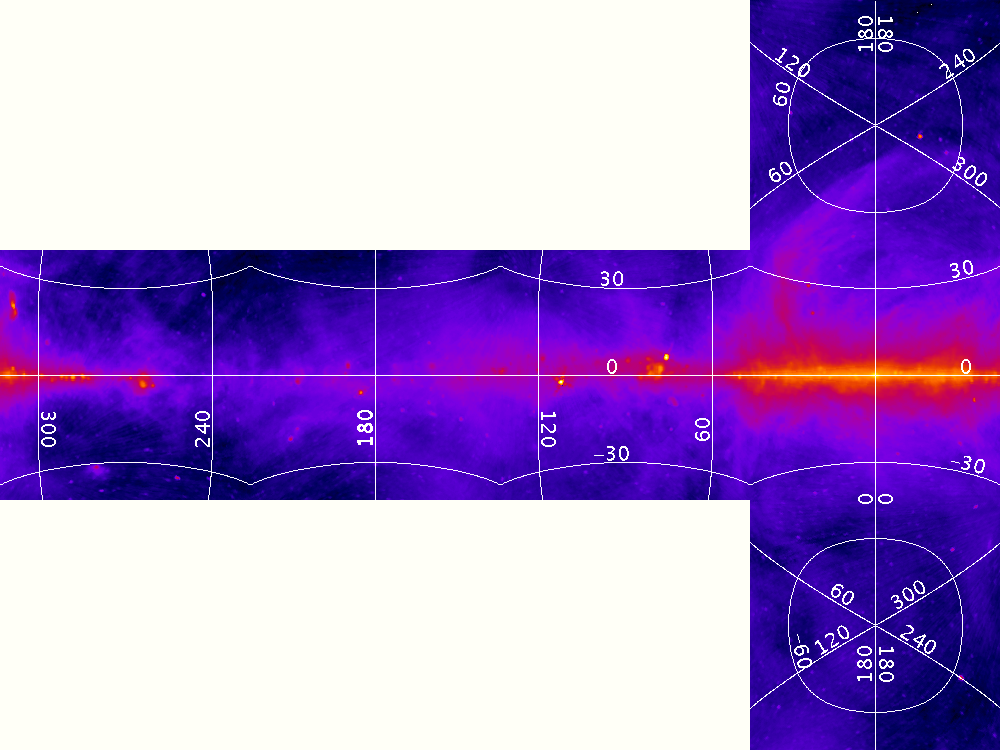}
\caption{Cubic quadralaterized projections.   The sphere is mapped to the six faces of the cube which may be combined into a single plane.  This and subsequent all-sky maps are reprojections by \textit{SkyView} rendering the \citet{Haslam1982} 408 MHz all sky map using the 'fire' color table and logarithmic brightness scaling. This image uses the  CSC projection. \label{fig:cubic-qaud-proj}}
\end{figure}

Determining the face that a given celestial position maps to is non-trivial when we work using coordinates, but can be done straightforwardly using the unit vectors on the sphere.  The face a given position maps to is determined by the index and sign of the unit vector component with the largest magnitude.  E.g., in Figure \ref{fig:cubic-qaud-proj}, the region in the sky where the z component of the unit vector is positive and larger than the other components maps to uppermost facet.   If the x component is largest and negative the point maps to the second facet from the left centered on the Galactic anticenter.

This approach enables us to determine the corner locations easily.  These are just the points at which the unit vector components are of equal magnitude.  Since their squares must sum to unity, each component must be $\pm \sqrt{3}/3$ .  Thus the latitude of the corners is $\sin^{-1}⁡ \sqrt{1/3}$ or about $35^{\circ}$.  The distance from the center of a tile (e.g., the pole) to a corner must be $\sin^{-1}⁡ \sqrt{2/3}$ or about $55^{\circ}$.

Once we have split the sky into face squares, they can be arranged in a variety of ways. The T shown in Figure \ref{fig:cubic-qaud-proj} is one example, but one can `roll' the polar facets to wherever is convenient.  Figure \ref{fig:cubic-qaud-proj} shows  that while the transformation is continuous at the cube boundaries, there is a very significant discontinuity in the derivatives there.  We see major kinks in the coordinate grid at facet edges. 

Another aspect of the cube projections is that they cannot be naturally extended to fill the projection plane.  Although we can extend either of the lines in the T indefinitely, we cannot fill in the space inside the corner of the T.  To do so a tile would need to have two identical edges.  

This discussion so far applies to all of the cube projections.  A complete projection needs some algorithm which takes the coordinates in a sixth of the sky and maps it to a square.  

A simple approach is to visualize embedding the unit sphere in a cube with the faces of the cube tangent to the sphere.  If we draw a line from the center of the sphere it passes through the sphere and then the cube mapping the position on the sky to a specific location on a specified cube face.  The sky is split into six tangent planes.  This is the Tangent Spherical Cube (TSC).

The two other cube projections included in \citet{Calabretta2002} use more complex transformation rules to reduce (CSC) or eliminate (QSC) variations in pixel area within the projection (see \citet{Chan1975, Oneill1976,1992ASPC...25..379W} for details). 
% * <tom.mcglynn@nasa.gov> 2018-05-04T14:36:48.545Z:
% 
% Added references per referees comment 4.
% 
% ^.

The approach we have taken below is quite similar to the cube projections, but is based upon the octahedron rather than the cube.  In the next section we discuss the overall approach to the transformation.  The following section describes three specific implementations of these: a tangent plane approach similar to the TSC projection, an equal area approach which has significant similarities to the HEALPix \citep{Gorski2005} projection and is similar in motivation to the QSC cube projection, and a projection based on recursive averaging of vectors.  This last approach, the Tessellated Octahedral Adaptive Spherical Transformation  or TOAST is particularly easy to compute on hierarchical tiles and is used in the WorldWide Telescope\footnote{http://www.worldwidetelescope.org} \citep{Rosenfield2018}.  We expound upon the properties of the TOAST projection in Section 5 and discuss the differences between the TOAST projection and the TOAST pixelization.

\section{A topological approach based on the octahedron}
While the cube projections did not meet our original criteria the cube is not the only solid we can embed a sphere inside.  If we use an octahedron, a simple rescaling enables us to generate projections that meet all of our desired criteria. 

\begin{figure}
\plotone{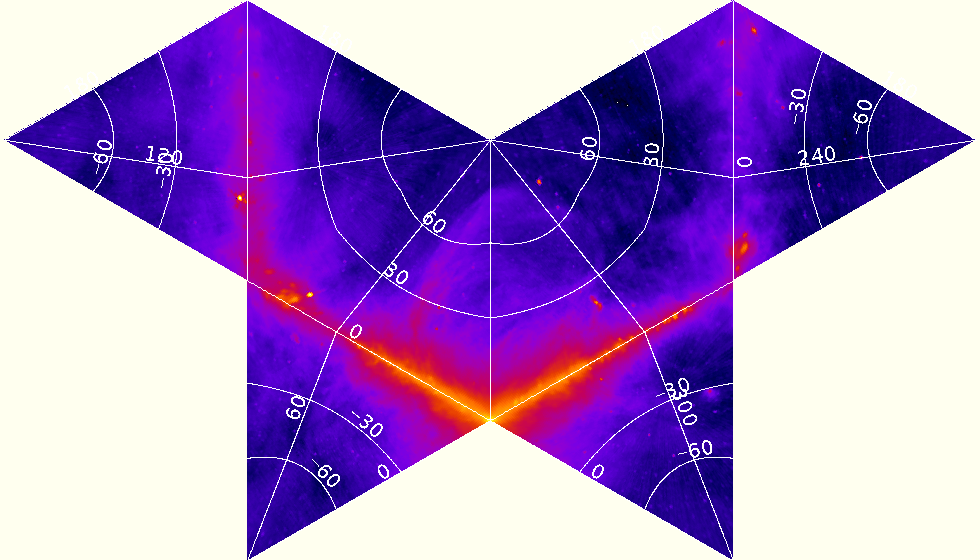}
\caption{Octahedral projection using a tangent plane on each face. (SkyView image using the internally defined CAH projection). \label{fig:CAH-proj}}
\end{figure}

Using octahedrons is not new.  An early use of an octahedral decomposition of the globe was Cahill's Butterfly World Map \citep{Cahill1909} a rounded version of the projection shown in Figure \ref{fig:CAH-proj}.

An octahedron is 8 equilateral triangles arranged as two pyramids pointing away from each other.  Cahill noted that the octahedron can be unwrapped into an elegant butterfly shape which can be arranged for an Earth projection such that the only continent significantly chopped up is Antarctica.  Even this may be less problematic then the omission of most of the continent in traditional Mercator maps. In Figure \ref{fig:CAH-proj} we show a variant of Cahill's projection where each of the triangles is a tangent plane projection of one octant of the sky (using the same survey as Figure 1).  While this shape may be attractive, the non-convex nature violates one of our desired characteristics - that the map be a simple rectangle.

\begin{figure}
\plotone{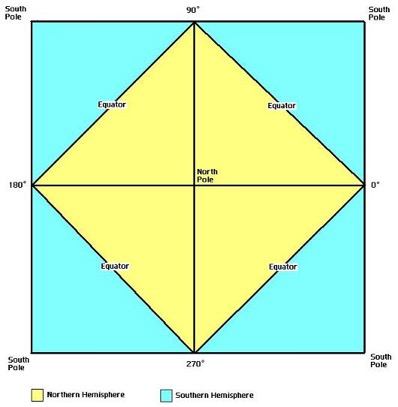}
\caption{Transforming the octahedron to a square.  Looking from directly above the octahedron one would see the northern facets of the octahedron projected to right triangles.  The southern facets swing out on the equator to transform the diamond to a square. \label{fig:transform-octah-sq}}
\end{figure}

To get a rectangular map from an octahedral projection, we transform the equilateral triangles into right isosceles triangles.  One simple transformation squashes each triangle down from the pole (or equivalently stretches it along the base at the equator).  E.g., we can look down at the octahedron from directly above.   The viewer will see the facets of the octahedron appear as right triangles forming a diamond or square as shown in Figure \ref{fig:transform-octah-sq}.  The northern facets conceal the southern facets below them.  Given 8 right triangles there are many possible arrangements that we can make of the triangles to form a rectangle.  We choose to use the equator lines as hinges, and flip out the southern facets  to form a square centered on the North Pole with the South Pole split to each of the four corners of the square.

Alternatively, one can envisage squashing the triangles in Figure \ref{fig:CAH-proj}, pressing down from the poles.  As the northern hemisphere triangles flatten, the gap that splits the north at 180$^\circ$ closes.  When it closes we are left with the same shape as of Figure \ref{fig:transform-octah-sq}.

We now have a general prescription for building a projection that meets all of our criteria so long as the -- as yet unspecified -- transformation between an octant of the sphere and the facet of the octahedron is singularity free.  Since we only need to transform an eighth of the sky for each facet avoiding singularities is not difficult.

Finding the facet that particular coordinates in the sky map to is easier when we use an octahedron compared to the cube projections.  Since we are mapping octants, the equator divides facets in latitude, and the facets divide in longitude in 90 degree segments.  If we deal with unit vectors then we can use just the signs of the unit vector components to map to the facet.  

Since each triangle is stretched, we anticipate that our projection should be continuous at the boundaries where we joined tiles, but it is unlikely to be differentiable there, since the stretching of different facets is in a different direction.   This is similar to what we saw with the cube projections.

While the sphere projects to a finite square in the projection plane in this approach, the entire projection plane can be covered by tiling the plane with replications of the central square.  Adjacent tiles are rotated by 180 degrees.  There are no hard boundaries to the projection.  The projection plane can be covered with interleaved diamonds of the two hemispheres. 

If we recall our impetus for these projections, to support computer processing, the octahedral-based approach has several significant advantages compared to the cube-based projections:  our key gain is the ability to represent the sky in a single rectangle, but we also find the mapping between sky and facets is easier and that there are no real edges to the projection.

\section{Specific projections}
In discussing the transformation from sky to plane, we need only discuss one octant.  If we understand the projection in the prime octant where the longitude, $\alpha$, has $0 \le \alpha \le 90^\circ$ and the latitude, $\delta$, has $0\le \delta \le 90^\circ$, then by symmetry we can compute the projection for all of the other octants. 

Suppose that we have some function $(x_0,y_0) = f(\alpha_0,\delta_0)=g(u_0,v_0,w_0)$ defined over coordinates, $(\alpha_0,\delta_0)$, or a unit vector, $(u_0,v_0,w_0)$, in the prime octant.  Then anywhere on the sphere we may transform from $(\alpha,\delta)$ to $(\alpha_0,\delta_0)$ or from $(u,v,w)$ to $(u_0,v_0,w_0)$  and then transform the resulting $(x_0,y_0)$ back to $(x,y)$.  Table \ref{table:transforms} describes one set of consistent transformations where we scale the transformation such that we are filling the square in the projection plane with $|x|, |y| \le s$ with a total area in the plane of $4s^2$.  The first column numbers the octants.  The next two define the octant in terms of coordinate ranges or signs of the unit vector components.  The fourth and fifth columns indicate how we transform the actual coordinates or unit vectors to the corresponding values in the prime octant.  The last column indicates how we transform the projected position we get from our function to the actual location in the projection plane.  It includes the scaling parameter, $s$, that depends upon the scaling we choose for the projection.
% * <tom.mcglynn@nasa.gov> 2018-04-27T20:00:35.844Z:
%
% ^.

\begin{deluxetable}{llllll}
\tablecaption{Transformations to and from the prime octant \label{table:transforms}}
\tablehead{
\colhead{Octant} &
\colhead{Coordinates} &
\colhead{Unit Vector} &
\colhead{Coordinate} &
\colhead{Unit Vector} &
\colhead{Projected} \\
\colhead{} &
\colhead{} &
\colhead{Signature} &
\colhead{Transform} &
\colhead{Transform} &
\colhead{coordinates} \\
\colhead{} &
\colhead{} &
\colhead{} &
\colhead{$(\alpha_0, \delta_0)$} &
\colhead{$(u_0, v_0, w_0)$} &
\colhead{$(x, y)$} \\
}
\startdata
0	&	$0   \le \alpha \le  90^\circ$	&	$+ + +$	&	$(\alpha,     \delta)$	&	$( u, v, w)$	&	$(x_0, y_0)$ \\
	&	$0   \le \delta \le  90^\circ$	&   		&	                    	&	            	&	\\
1	&	$0   \le \alpha \le  90^\circ$	&	$+ + -$	&	$(\alpha,    -\delta)$	&	$( u, v,-w)$	&	$(s-y_0, s-x_0)$ \\
	&	$-90 \le \delta \le   0^\circ$	&  	    	&	                    	&	            	&	\\
2	&	$90  \le \alpha \le 180^\circ$	&	$- + +$	&	$(\alpha-90,  \delta)$	&	$( v,-u, w)$	&	$(-y_0,x_0)$ \\
	&	$0   \le \delta \le  90^\circ$	&  		    &	                    	&	            	&	\\
3	&	$90  \le \alpha \le 180^\circ$	&	$- + -$	&	$(\alpha-90, -\delta)$	&	$( v,-u,-w)$	&	$(x_0-s,s-y_0)$ \\
	&	$-90 \le \delta \le   0^\circ$	&  	    	&	                    	&	            	&	\\
4	&	$180 \le \alpha \le 270^\circ$	&	$- - +$	&	$(\alpha-180, \delta)$	&	$(-u,-v, w)$	&	$(-x_0,-y_0)$ \\
	&	$0   \le \delta \le  90^\circ$	&  		    &	                    	&	            	&	\\
5	&	$180 \le \alpha \le 270^\circ$	&	$- - -$	&	$(\alpha-180,-\delta)$	&	$(-u,-v,-w)$	&	$(y_0-s, x_0-s)$ \\
	&	$-90 \le \delta \le   0^\circ$	&  		    &	                    	&	            	&	\\
6	&	$270 \le \alpha \le 360^\circ$	&	$+ - +$	&	$(\alpha-270, \delta)$	&	$(-v, u, w)$	&	$(y_0,-x_0)$ \\
	&	$0   \le \delta \le  90^\circ$	&  	     	&	                    	&	            	&	\\
7	&	$270 \le \alpha \le 360^\circ$	&	$+ - -$	&	$(\alpha-270,-\delta)$	&	$(-v, u,-w)$	&	$(s-x_0,y_0-s)$ \\
	&	$-90 \le \delta \le   0^\circ$	&  		    &	                    	&	            	&	\\
 \enddata
 \end{deluxetable}
 
\subsection{Triangular octahedral tangent plane (TOT)}
 One simple projection corresponds directly to TSC, the tangential spherical cube projection.  We just embed the sphere inside octahedron such that sphere is tangent to each facet.  We draw lines from the center of the sphere through the sphere and octahedron, mapping each position on the sphere to a unique facet and position in the polyhedron.
 
The tangent plane projection around an arbitrary point can be represented as  
\begin{eqnarray}
x &=&  \frac{\cos⁡ \delta  \sin⁡(\alpha-\alpha_0 )} {\cos⁡ c} \\
y &=&  \frac{ \cos⁡\delta_0 \sin⁡\delta - \sin⁡\delta_0 \cos⁡\delta  \cos⁡(\alpha-\alpha_0)}{\cos⁡c}
\end{eqnarray}
where
\begin{equation}
\cos⁡c=  \sin⁡\delta_0 \sin⁡\delta+  \cos⁡\delta_0\cos⁡\delta\cos⁡(\alpha-\alpha_0)
\end{equation}
and $(x,y)$ is the point corresponding to the right ascension and declination $(\alpha, v)$.  The tangent point of the projection is $(\alpha_0, \delta_0)$.  Here $c$ is just the angular distance to the reference point for the projection.

The inverse projection from the tangent plane to the sphere is
\begin{eqnarray}
\delta &=& \sin^{-1}⁡\left(\cos⁡c\sin⁡\delta_0+ \frac{y \sin⁡c  \cos⁡\delta_0}{p}\right) \\
\alpha &=& \tan^{-1}⁡\frac{x \sin⁡c}{p \cos⁡\delta_0  \cos⁡c-y \sin⁡\delta_0 \sin⁡c}
\end{eqnarray}
where
\begin{eqnarray}
p= \sqrt{x^2+y^2}
c=\tan^{-1}⁡p  
\end{eqnarray}
Here $p$ is the distance in the projection plane from the reference location, while $c$ is again the distance on the celestial sphere.

If the reference point is set to the pole, i.e., $(\alpha_0, \delta_0)$= $(0,90^\circ)$ then we have a much simpler transformation:
\begin{eqnarray}
x&=&\frac{\sin⁡\alpha}{\tan⁡\delta} \\
y&=&\frac{\cos⁡\alpha}{\tan⁡\delta}
\end{eqnarray}
With the inverse
\begin{eqnarray}
\alpha&=&\tan^{-1}⁡(y,x) \\
\delta&=&\cot^{-1}⁡(x^2+y^2)
\end{eqnarray}

In practice one can often rotate the center of the tile to the pole and use the simpler equations.  This rotation may be combined with any rotation needed to move to the prime octant.

By symmetry the tangent point for each facet when we enscribe a sphere inside an octahedron must be equidistant from each of the three vertices of the facet.   We can find this as the point of the intersection of the angle bisectors for the spherical triangle defining the octant.

\begin{figure}
\plotone{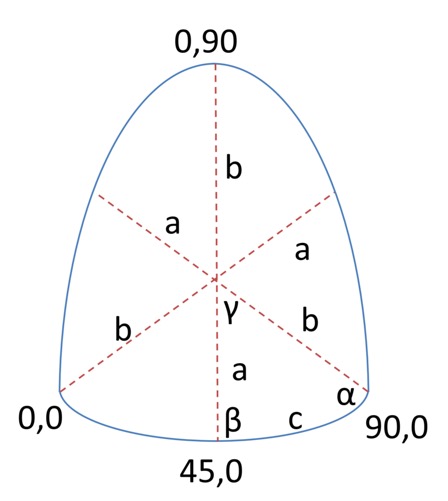}
\caption{Finding the center of the octant.\label{fig:center-octant}}
\end{figure}

If we have oriented the octahedron with vertices at the poles and coordinate origin as shown, then we might see the prime octant in Figure \ref{fig:center-octant}. Here the blue lines represent the boundaries of the octant, a spherical triangle with vertices at (0,0), (90,0) and (0,90).  The red lines are the angle bisectors.  The center of the triangle is at the point where the bisectors meet. The angle bisectors also bisect the opposite sides. One of the bisectors is simply the meridian at $45^\circ$ longitude.  This gives the longitude of the center, while the latitude is clearly just the angle, $a$, along that meridian.

If we look at the triangle with angles $(\alpha,\beta,\gamma)$ we note that $\alpha$ is the result of bisecting the 90 degree angle between the prime meridian and the equator, so $\alpha=45^\circ$.  The angle $\beta$ is the angle between the meridian at $45^\circ$ and the equator, so $\beta=90^\circ$.  Finally by symmetry all of the angles where the bisectors meet must be equal (since we started with an equilateral triangle), so $6\gamma = 360^\circ$, or $\gamma=60^\circ$.     By the law of sines for spherical triangles we have  $\frac{\sin⁡a}{\sin⁡\alpha} =   \frac{\sin⁡b}{\sin⁡\beta} =   \frac{\sin⁡c}{\sin⁡\gamma} $.  We know all of the angles, and $c=45^\circ$.  So $\frac{\sin⁡a}{\sin⁡45^\circ} =  \frac{\sin⁡45^\circ}{\sin⁡60^\circ}$ .  Hence $\sin a=1/\sqrt{3} $ or $a \approx 35.264389^\circ$.

The distance between the center of the facet and the corners, $b$ in Figure \ref{fig:center-octant}, has $\sin⁡b=\sqrt{2/3}$ so $b \approx55^\circ$.   Despite there being two more facets than in the cube projections we have the same maximum distance as with the TSC projection.  Squares are more efficient in packing data closer to the center.  This may be more intuitive if we note that a cube has 8 vertices compared to the octahedron's 6.  

The tangent plane projection is radially distorted with projected radius growing as the tangent of the actual radius.  The distortion, (the ratio of the apparent change in radius to the actual change) is just the derivative of the tangent.  At the corners this is exactly 3.  However, a relatively smaller fraction of the facets are in the regions of highest distortion for the octahedron compared to the cube.

In this projection (see Figure 5), straight lines correspond to great circles, but there can be a kink where we cross octant boundaries.

There are multiple scales which may be appropriate for the projection.  When we project an octant onto the projection plane we get an equilateral triangle with sides of $\sqrt{6}$, or an area of $3 \sqrt{3}/2$.  Thus the total projected area in all eight triangles is $12\sqrt{3}$.  If we wish to preserve this area, then our projection should be a square with sides $2 (3^\frac{3}{4})$ in radians or about $261.21^\circ$.  With this scaling the projection should conserve area near the center points of each tile.  However the Tissot Indicatrices are still elliptical due to the squashing of the triangles.

Alternatively, we could include the squashing of the triangles as part of the scaling of the projection.  The area of each triangle is reduced by a factor of $1/\sqrt{3}$ as we squash from an equilateral to an isosceles right triangle.  This gives us a total area of 12 or a side of $2\sqrt{3}$ radians or about $198.478^\circ$.  This area is very close to $4\pi$, so that the average areal distortion over the map is small, the expansion due to the projection almost exactly compensating for the squashing of the equilateral triangles.    This does not make the map any less distorted, the projection shrinks some regions as it expands others.

Using this scaling we can calculate $f(\alpha_0,\delta_0)$ for the prime octant.   First we project onto the plane using the tangent plane projection around the tile center.  We shift the resulting equilateral triangle, squash it, and rotate it into position.

Let $f_p (\alpha,\delta)=(x_p,y_p )$ be the results of projecting a position in the prime octant to the tangent plane centered on the center of the octant. We have
\begin{eqnarray}
x_p&=& \frac{\sqrt{3}  \cos⁡\delta  \sin⁡(\alpha-45^\circ)}{\sin⁡\delta+\sqrt{2}  \cos⁡\delta \cos⁡(\alpha-45^\circ)}\\
y_p&=&\frac{\sqrt{2}  \sin⁡\delta-\cos⁡\delta  \cos⁡(\alpha-45^\circ)}{\sin⁡\delta+\sqrt{2}  \cos⁡\delta  \cos⁡(\alpha-45^\circ)}
\end{eqnarray}

\begin{figure}
\plotone{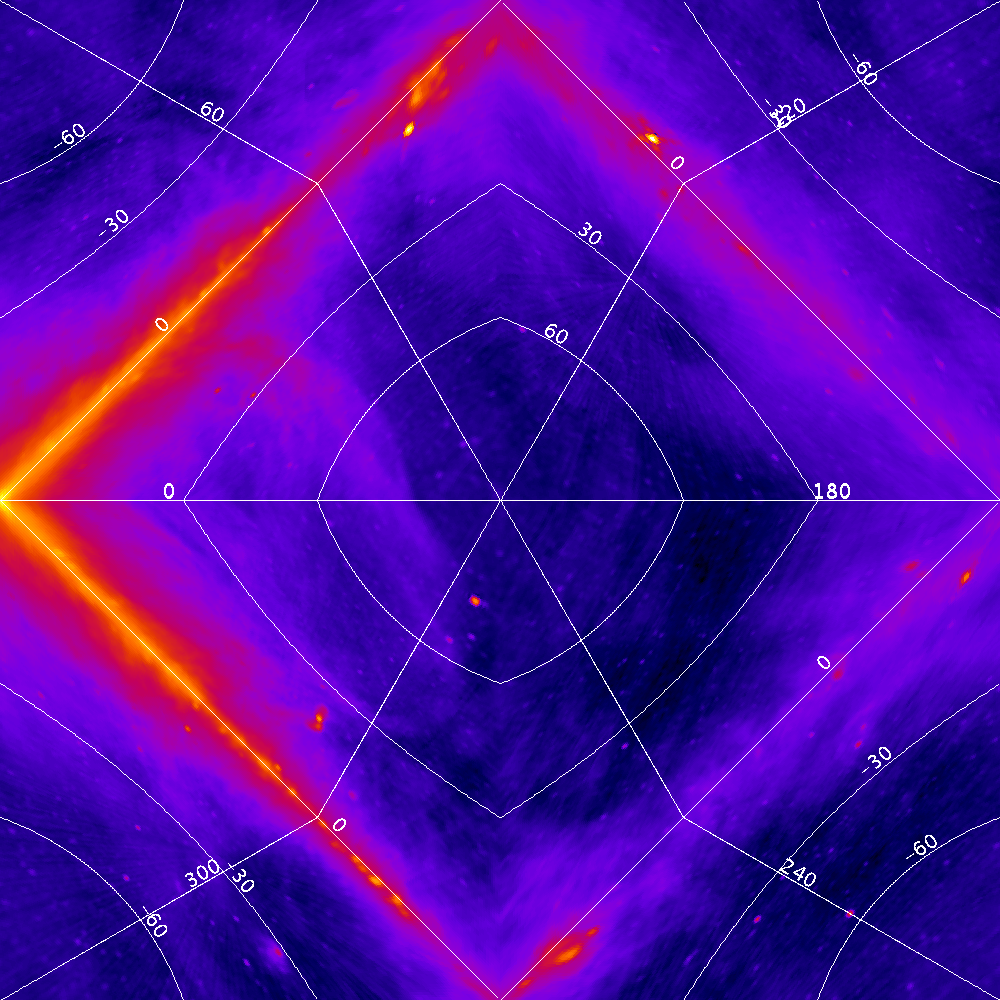}
\caption{The Triangular Octahedral Tangent  Projection (TOT).\label{fig:TOT}}
\end{figure}

When we apply these equations to  projects the prime octant we get  an equilateral triangle centered on the origin with the pole at $(0,\sqrt{2})$.  We must shift the pole to the origin, rescale the triangle to an isosceles right triangle with the right angle at the pole vertex and then rotate the plane coordinates to the appropriate orientation to match its location in Figure 3. So
\begin{equation}
\mathbf{f}(\alpha,\delta)= \mathbf{R} \mathbf{S} [\mathbf{f_p} (\alpha,\delta)-\mathbf{T}] 
\end{equation}

Where $\mathbf{S}=\left[\begin{array}{cc}1&0\\0&(1/\sqrt{3})\end{array}\right]$ does the squashing, $\mathbf{T}= \left[\begin{array}{c}0 \\ \sqrt{2}\end{array}\right]$ is the translation vector and $\mathbf{R}=\left[\begin{array}{cc} -\frac{1}{\sqrt{2}} & -\frac{1}{\sqrt{2}} \\ \frac{1}{\sqrt{2}} & -\frac{1}{\sqrt{2}}\end{array}\right]$ does a rotation by 135$^\circ$ to get it to the appropriate location in the coordinate tile illustrated in Figure 3.     The formulae in Table \ref{table:transforms} show how we can address the other octants with the scaling parameter $s=\sqrt{3}$, i.e. the entire sky will be represented in the projection plane with $x,y$ values between $\pm\sqrt{3}$.

\subsection{An octahedral equal-area projection}
It is straightforward to define an equal area projection for each tile.  E.g., if we make the parallels of latitude lines in the projection plane, then the constraint that the area north of a given latitude must be preserved defines the projection.  The transformation $f(\alpha,\delta)=$(x,y) for the prime octant is just
\begin{eqnarray}
t&=& 2\sqrt{\frac{1-\sin⁡\delta}{2}} \\
u&=& \frac{2}{\pi}  t \alpha \\
x&=& \sqrt{\frac{\pi}{2}} (t-u) \\
y&=& \sqrt{\frac{\pi}{2}}  u 
\end{eqnarray}
The intermediate values $(t,u)$ range from 0 to 1 (in the prime octant).  This projection maps directly to a right isosceles triangle in the appropriate orientation. For other octants we use the transformations in Table \ref{table:transforms} with a scale factor of $\sqrt{\pi}$.  The inverse transformation is given by
\begin{eqnarray}
t&=&\sqrt{\frac{2}{\pi}}(x-y)\\
u&=&\sqrt{\frac{2}{\pi}}y\\
\delta&=&  \sin^{-1}⁡\left(1-\frac{t^2}{2}\right)\\
\alpha&=&  \frac{\pi}{2} \frac{u}{t} 
\end{eqnarray}

This is an exact, equal area projection which we have called the Triangular octahedral Equal Area or TEA projection.  This is essentially the Collignon projection over the area of each of each facet \citep[see][]{Calabretta2007}. 

The TEA projection is closely related to the map projection for HEALPix \citep{Gorski2005}.  Since we have constructed our projection to have the latitudes in straight lines, it is not surprising that the projection is equivalent to another equal area projection with latitudes defined to be constant in one dimension. The HEALPix projection can be rendered as map with a sawtooth triangles for latitudes $| \delta | \gtrsim 50^\circ$.  These correspond to the tips of the triangles in the TEA projection.  However, the TEA projection does not transition to a different projection at lower latitudes as HEALPix does.  At lower latitudes, HEALPix transitions to a cylindrical equal area projection which leads to less distortion among the pixels but means that there are different classes of pixels in HEALPix.

\begin{figure}
\plotone{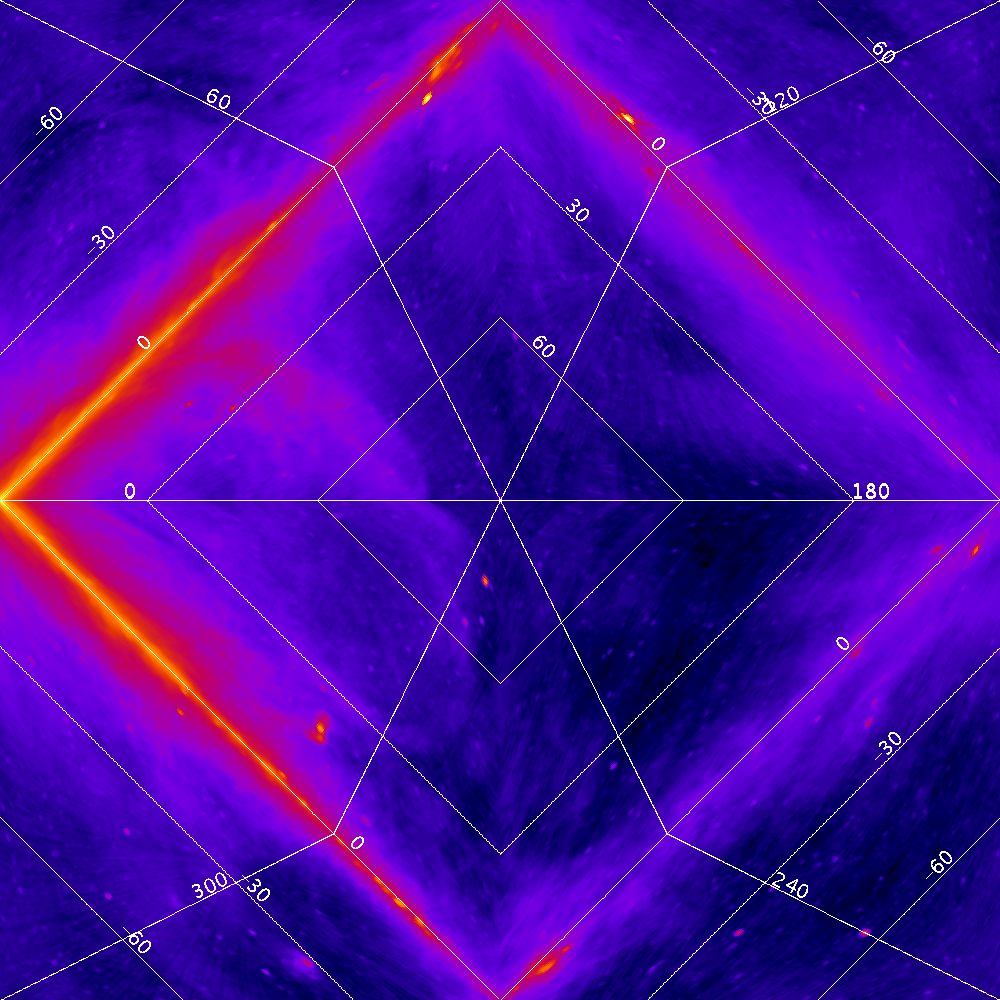}
\caption{The Triangular Octahedral Equal Area  Projection (TEA).\label{fig:TEA}}
\end{figure}

Since the TEA projection (Figure \ref{fig:TEA}) is an equal area projection, an appropriate scale for the projection is for the standard projection region to cover an area of $4\pi$ when we are working in radians.  This gives an edge dimension of $2\sqrt{\pi}$ in radians.  This corresponds to about 203.10825$^\circ$.  The factors of $\sqrt{\pi/2}$ in the forward projection above set this scale.

A  Collignon projection of the full sphere can be rendered with the hemispheres transformed to triangles sharing a base at the equator to create a single all-sky diamond-shaped tile.    It is straightforward to scale the diamond to a square which then meets all of our requirements for a desireable projection: all-sky, rectangular, no singularities and extensible.  To tile the projection plane, four copies of the all-sky tile meet at each pole.  Near the poles there will be four points corresponding to the same point on the sky.  In the TEA projection (and the other projections using this topology), there can be 2 nearby points corresponding to the same point in the sky  where two diamonds representing the same hemisphere touch.

\subsection{The tessellated octahedral adaptive spherical transformation}

The Tessellated Octahedral Adaptive Spherical Transformation (TOAST) does not use an analytic transformation between the celestial sphere and projection plane.  Rather it uses a hierarchical transformation identical to those of the Hierarchical Triangular Mesh (HTM) pixelization popularized by the Sloan Digital Sky Survey \citep[see][]{Kunszt2001,Kazdhan2010}.

Consider the vertices of one of the facets of the octahedron, which we now take as being inscribed inside the sphere, so that the vertices are on the surface of the sphere.   Take the arcs between the three points and find the midpoint (on the sphere) between them.  E.g., if we label the initial points A,B,C, we find the midpoints of AB, BC and CA (call them x, y and z respectively).  Note that we used the arcs between the points so that x, y and z are also on the surface of the sphere.   We may now subdivide the original triangle into four sub-triangles  Axz, Byx, Czy and xyz. We can repeat this process recursively until we have as dense a mesh of points on the sphere as we desire.  

\begin{figure}
\plotone{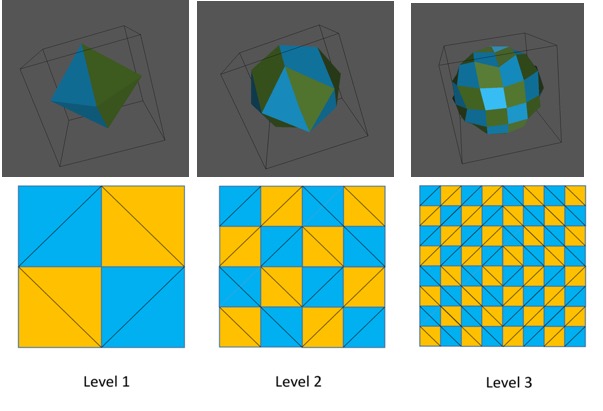}
\caption{Transforming the sphere to the plane in TOAST.  The top row shows successive subdivision of the sphere while the bottom gives the corresponding triangulation of the projection plane.  Note that in the Level 1 image in three dimensions we show all 8 facets of the tetrahedron even though they are paired in the TOAST pixels –- as shown in the coloring. \label{fig:trans-TOAST}}
\end{figure}

Corresponding to the vertices on the sphere, we define points in the projection plane.  We start with the eight right-triangles corresponding to the facets of the octahedron.  As we subdivide the triangles on the sphere, we correspondingly subdivide the triangles on the plane and identify the corresponding new vertices.   Figure \ref{fig:trans-TOAST} shows this correspondence.  We obtain a finer and finer mesh on the sphere and a corresponding mesh in the projection plane. 

We may define the recursion more rigorously as:
\begin{enumerate}
\item	Consider three points $\mathbf{p_0,p_x,p_y}$ in the projection plane at $(x,y)$, $(x+\delta, y)$, and $(x,y+\delta)$.  These have already been defined as associated with three unit vectors $\mathbf{u_0,u_1,u_2}$ on unit sphere.  We start the recursion with $(x,y) = (0,0), \delta=1, \mathbf{u_0}=(0,0,1), \mathbf{u_1}=(1,0,0), \mathbf{u_2}=(0,1,0)$.
\item Associate new points $\mathbf{p_{0x}}$, $\mathbf{p_{0y}}$ and $\mathbf{p_{xy}}$ with the unit vectors $\mathbf{u_{01}}$, $\mathbf{u_{02}}$, and $\mathbf{u_{12}}$ where $\mathbf{p_{0x}} = (x+\delta/2, y), \mathbf{p_{0y}}=(x,y+\delta/2)$, and $\mathbf{p_{xy}}=(x+ \delta/2, y+\delta/2)$ and $\mathbf{u_{01}}=\frac{\mathbf{u_0}+\mathbf{u_1}}{|\mathbf{u_0}+\mathbf{u_1}|}$ and similarly for $\mathbf{u_{02}}$ and $\mathbf{u_{12}}$.
\item If we need finer resolution than so far achieved, recurse in each of four subtriangles using ($\mathbf{p_{0y}}$, $\mathbf{p_{xy}}$, $\mathbf{p_y}$), ($\mathbf{p_0}$, $\mathbf{p_{0x}}$, $\mathbf{p_{0y}}$), ($\mathbf{p_{0x}}$, $\mathbf{p_x}$, $\mathbf{p_{xy}}$) and ($\mathbf{p_{xy}}$, $\mathbf{p_{0y}}$, $\mathbf{p_{0x}}$) as the entry points in step 1.  At each level the magnitude of $\delta$ will halve.  For the last triangle, the sign of $\delta$ is inverted.
\end{enumerate}
To cover the entire sky we can use the transformation approach described in  Table \ref{table:transforms}  to handle positions outside the prime octant, or we can start with 8 triangles directly using the appropriate points and unit vectors.

The TOAST projection is explicitly defined only at the grid points associated with the HTM recursion.  These constitute a set of measure 0 in the projection plane (or on the celestial sphere).  Looking only at these grid points the transformation is continuous: as we consider grid points that differ by smaller and smaller amounts, the values of the grid points converge.  Since the grid can be made arbitrarily dense, we can extend the transformation to an arbitrary point by \textit{defining} the transformation to be continuous.  I.e., to find the value of the transformation at some arbitrary point not on the grid, we look at grid points sufficiently nearby to satisfy whatever precision requirements we have. 
% * <tom.mcglynn@nasa.gov> 2018-04-18T21:18:09.896Z:
%
% ^.
Note that triangles on the sphere are not of identical size.  Even though we start with equilateral triangles, the inner triangle (the last one in step 3) is different from the other three.  However the variation in size of triangles is rigidly bounded.  All triangles of a given level of the recursion in the projection plane are of the same size by construction. Thus TOAST is not an equal area projection.  In fact, since we create different size triangles at each step, the transformation between the sky and the plane is not differentiable at the points it is calculated although it is continuous.  At the largest scales-- at the boundaries between the top-level triangles – these distortions can be easily visible but  becomes  less obvious at smaller scales (see Figure 9).

We can use the recursion to build grids of pixels in TOAST.  The level 0 pixel is the entire sky.  At level 1, each of the four pixels covers  90 degrees of longitude.   At the next level, these longitude bands are each split into four pixels, two of which touch the poles and two of which straddle the equator.  At each level we have a $2^n \times 2^n$ grid.

\begin{figure}
\plotone{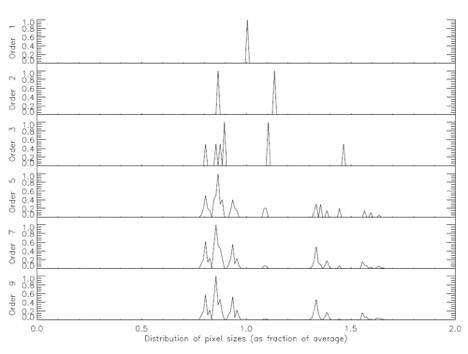}
\caption{Distribution of sizes of TOAST pixels.  The distribution of the sizes of pixels of a given order relative to the average size of pixels of that order.  The distribution stabilizes as we go to higher order.\label{fig:toast-px}}
\end{figure}

The computation of the TOAST/HTM boundaries is discussed in detail in \citet{Szalay2005}.  Figure \ref{fig:toast-px} is comparable to their Figure 3 and gives the distribution of pixel sizes as a fraction of the average size for a variety of values of n where we represent the entire sky as a $2^n \times 2^n$ grid.  One slight difference with Szalay et al. is that we are giving the sizes of pixels which comprise two adjacent triangles rather than individual triangles but the essence of the figure is unchanged.  All level one pixels have the same size.  We see that level 2 pixels come in two distinct sizes.  As we move towards higher values of $n$ the histogram stabilizes into a complex pattern. This form of the histogram takes a few levels to build up from the congruent level 1 pixels.  Thus globally the TOAST projection will appear less distorted than the histograms suggest.  At a given level pixels may differ in area by a maximum factor of about 2.
% * <tom.mcglynn@nasa.gov> 2018-04-18T20:41:42.658Z:
%
% ^.

When we work using unit vectors, the TOAST recursion is particularly simple.  The initial conditions for the recursion are trivial and the recursion itself is very fast.  The cost of this computation is essentially the computation of a single square root.  This computation is simple enough that it may be done in graphics processors as well as a computer's main CPU.

In the typical case where we start with a top level tile of at least 256x256 pixels,  calculation of the square root can be simplified.  This is used to get the length of the average of two unit vectors -- which is just the cosine of the angle between them.  Except for the top level tile we are generally dealing with angles less than 1 degree, so that we will only need to deal with square roots in a very small range 0.999 - 1 which can potentially be calculated using a simpler and more efficient algorithm than the general function over all positive real numbers.

Since we start with the entire sky and recurse by factors of two into increasingly fine tiles, the TOAST projection provides a very rapid way to find the positions of a hierarchical image tiles when we define the tiles to match the recursion, i.e., the tiles should comprise subsets of the $2^n \times 2^n$  all-sky grids.  If  we have divided the sky into a grid of pixels for some value of n, then a natural tiling is to divide the sky into a grid of $2^k \times 2^k$ tiles where each tile has $2^m \times 2^m$ pixels with $n=m+k$.  The positions of pixel corners in these standard tiles can be computed at a cost of roughly a single square root per pixel.  The small number of recursions needed to find the bounding box for the tile is amortized over the many pixels within the tile.

While the transformation is easy to calculate for grids that match the recursion, the TOAST projection is less straightforward generally.  For an arbitrary position, we must use recursion to refine the projected position for a given set of coordinates.  We transform our position to the prime octant as defined in Table \ref{table:transforms}.  Then we recursively split this triangle in sub-triangles always finding the triangle our desired position is in.  

To determine this, we use a standard approach where we generate the three vectors which are the cross-products of pairs of the unit vectors to the vertices of the triangle.   We take adjacent vertices in a counterclockwise direction.  Since the cross-product is perpendicular to the plane containing the two unit vectors, it is perpendicular to the great circle connecting them.  With vertices chosen counterclockwise, the cross-products point into the triangle.  The dot product of our position unit vector with all three cross-products must be positive if the point is inside the triangle.  If the position is outside the triangle at least one of the dot products will be negative.  (This is not true in the general case where we would need to worry about spherical triangles with sides greater than 180 degrees, but our largest triangles are the octants whose sides are only 90 degrees.)

In practice roundoff errors can occasionally cause positions exactly on the boundary between two triangles to be seen as outside both of them.  In the case where a point seems be outside all four candidate sub-triangles, we use the triangle where the minimum of the three dot products for that triangle (which is negative since the point was found to be apparently outside the triangle), has the smallest magnitude.    

As we do the recursion on the sphere we update the position in the projection plane depending upon which subtriangle the position falls inside in this level of recursion.  Since each level of recursion reduces the area of the triangles by a factor of four, it doubles the positional accuracy.  To get a precision of $1\arcsec$ we can expect to need about 18 levels of recursion, or 28 levels of recursion to match within $0.\arcsec001$.    This implies that we need to go much deeper in the recursion to get comparable accuracy for an arbitrary grid compared to the natural tiles discussed above.  E.g., when we use the standard tiles we can go down to level 18 and we immediately have pixel corners specified to the limits of our arithmetic precision.  This would require going to a recursion depth comparable to the number of bits in the mantissa to duplicate for an aribtrary grid.  It is also more difficult to avoid recomputation  of higher level pixels for an aribitrary grid since we cannot be sure when we cross the boundaries of the natural pixels easily.  Thus the computation of an arbitrary grid within the TOAST projection will be much slower with each point requiring dozens of recursions rather than the single computation per pixel required for standard tiles.

If we have an arbitrary position in the projection plane for which we want the celestial coordinates, i.e., we need to de-project from the plane back to celestial coordinates, we invert the process.  Since the triangular grid in the projection plane is regular, we can easily calculate at each level of the recursion which subtriangle will be needed, i.e., which contains the point we are trying to get the coordinates for.  We start with the corner coordinates for the appropriate octant.  Then we gradually refine the celestial coordinates by finding the vertices of the ever smaller bounding triangles until one of the points in the current triangle in the projection plane is close enough to requested point that we can terminate the recursion. We use the corresponding celestial coordinates.

Since there is no analytic transformation, the scale of the TOAST transformation is arbitrary. Instead we consider the size of the box we transform into.  Often it is convenient to fit the entire sky into the 2x2 square centered at the origin so that the bounds of the projection are $|x|,|y|\le1$.   Alternatively, sides with dimensions of $\pi$ gives an image with where the average scale over the image is close to 1 if we assume distances in radians.  This gives a square with sides of 180$^\circ$.  There is no clear natural choice.

\section{Characteristics of TOAST}

\begin{figure}
\plotone{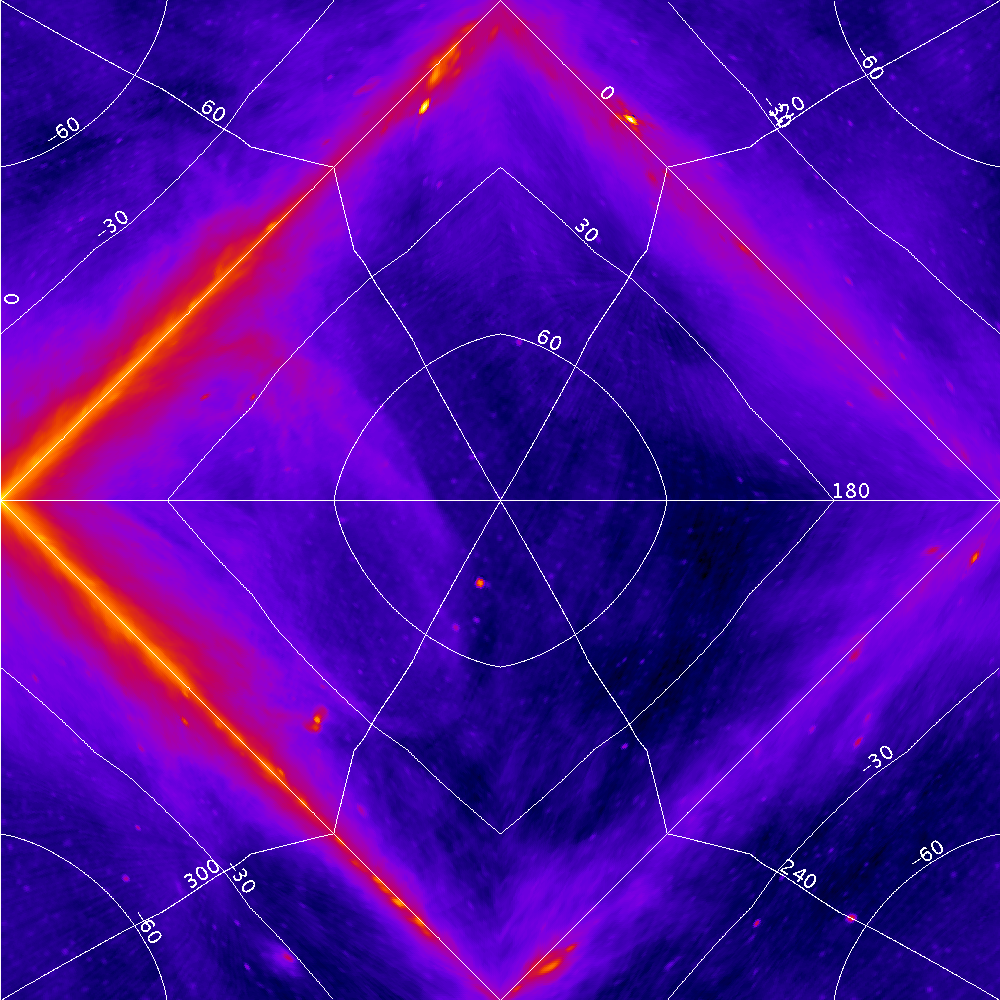}
\caption{The Tessellated Octahedral Adaptive Spherical Transformation (TOAST) Projection (TOA).\label{fig:TOA}}
\end{figure}

\subsection{Pixelization versus Projection}

It is useful to distinguish between the TOAST pixelization used in the WWT and elsewhere and the TOAST projection defined in this paper.  A projection is a transformation of points in the sky into some projection plane, a continuous transformation between sphere and a projection plane.  A pixelization is a specific scheme for dividing the sky into pixels with defined boundaries, essentially an integer valued function on the sphere or projection plane.  Often one defines a pixelization by imposing a rectangular grid over a projection plane.    This is not the case with TOAST.  The projection is defined in terms of a particular pixelization process which we extend beyond the grid points by assuming that the projection is continuous.  Thus, while it is possible in principle to define a pixelization of the TOAST projection plane different from the defining HTM pixelization, it is unlikely ever to be useful since the computation of the pixel edges will involve orders of magnitude more computation than for the fiducial HTM boundaries.  

The next section describes some of the characteristics of this particular pixelization that make it effective for the role of a machine intermediary representation of hierarchical resolution image data.
% * <tom.mcglynn@nasa.gov> 2018-04-27T20:11:24.399Z:
% 
% Deleted reference to section that we deleted
% 
% ^.

For the standard TOAST pixelization, the borders of each pixel are segments of great circles so that the boundaries are easy to compute.  However line segments other than pixel borders will not generally correspond to great circles, reflecting the special nature of the fiducial pixelization.

\subsection{TOAST and GPUs}

One of the motivations for using TOAST in the HTM pixelization is that it works efficiently with modern computers.   There are some terms specific to 3D accelerated graphics systems that we refer to here. For those not familiar with the field we start with a brief glossary of those terms. (For more detailed discussion see texts on image rendering, e.g., \citet{AkenineMoller2018}.)

 \textbf{Graphics Processing Unit (GPU)}.  A GPU is a specialized math co-processor, usually directly connected to an output frame buffer, that takes the input data in the form of small programs called shaders, lists of 2D or 3D graphics points and metadata known as vertex buffers, and graphics bitmaps known as textures and produces output as a rendered bitmap image. The shader programs run the same code against many input vertices in parallel thus they are much faster than graphics computed on the CPU.  This image is often directly displayed from the internal frame buffer to the user's display. GPUs are very common, and are often built-in to modern chips such as the Intel Core I5 and Core I7 chips used in Apple Mac and Windows PCs. These are also common on the System On Chip (SOC) components that make up tablets and cell phones.

\textbf{Texture}: A GPU based hosted image type made up of a pyramid of sub-sampled images. Textures are addressed by a set of normalized textual coordinates labeled U and V, collectively referred to as UV coordinates, with values between floating point values of 0 and 1. Textures are most often accessed through a texture sampler that uses a filter that may sample values in the surrounding region and at various levels of the pyramid to get the best representative value for the sample. Rarely does a sampler use just a single pixel value by itself. Samplers help prevent image aliasing that would occur in naively taking the sample from the nearest neighbor to a UV coordinate. GPUs typically draw images by interpolating the corners of a triangle in screen space and sampling textures using interpolated UV coordinates.

\textbf{Vertex Buffer}:  Vertex Buffers are a memory buffer hosted on the GPU that has a list of vertices and their associated metadata. By keeping this data in GPU memory, it is extremely fast to reuse and eliminates the need to specify the vertices to draw each frame.

\textbf{Index Buffer}: Index Buffers are a set of integer values that index into the Vertex buffer to define geometry, usually triangles. When vertices are used more than once, such as the corner of a cube that will be used by at least 6 different triangles needed to draw the cube, then it is wasteful to compute multiple copies of the vertex as it is transformed over and over. By using an index buffer the transformations can be computed once per vertex, and then used multiple times in drawing the various triangles that include that vertex by specifying geometry as a list of indices, rather than repeating the vertex values.

TOAST has some nice attributes that help it fill the need as an intermediate transmission format for large Internet delivered tiled-multi-resolution images rendered using GPUs.   The raw TOAST pixels are of very little use when displaying naively, but when transformed into the spherically projected environment by GPU rendering, tiles can be rendered efficiently and correctly projected.

The key attribute is the continuity of adjacency of pixels. For each tile level and across the TOAST projection, when UV texture maps are created with 3D meshes there are no discontinuities between adjacent pixels. Because GPUs use filters to sample pixels across multiple resolutions, when there are discontinuities in the input images, they cannot be used without the filter taking image data from unrelated areas of the sky for samples causing the output to be contaminated with unrelated data.

For example, some red pixels from a discontinuous part of the sky might be adjacent to blue pixels and the sampler would mix these two unrelated pixels to result in a violet color instead of red or blue. This would appear as visible artifacts at boundaries. Mitigating this would require more complex segmentation of images. Each discontinuous part of the sky would have to be broken up into different textures. This would reduce performance by requiring more separate draw operations on smaller vertex buffers and cause delays.  

Segmentation can also cause problems when panning the image.  Normally as a user pans across tile boundaries the program can defer to lower resolution tiles at some higher level in the tile hierarchy which will already be in memory since the antecedent lower resolution tiles are kept to hand.  The user gets an approximate image until the appropriate high detail tiles are retrieved. However if the image has multiple root segments then there may be no information available on the region being panned into and the display may stutter until the new root and higher resolution data can be retrieved.  

Because TOAST is continuous, images tiles from the top level down can be simply drawn as a batch with a single input image with no need to segment them.

Since TOAST is a fully determined projection with no free parameters (as opposed to something like a tangent plane projection, where the tangent point may vary), the mesh for any specific tile address is a deterministic set of 3D coordinates and UV coordinates regardless of the image content. This allows the use of pre-backed image meshes to be used either at tile creation time, or from a mesh cache.

Additionally, as tiled multi-resolution tile pyramids are generally drawn from the top down, each decedent tile can be calculated from their parent tile by simple mathematical subdivision. This allows meshes to be computed either in the GPU or CPU efficiently with simple and efficient algorithms.   In a segmented projection, this code would normally need to determine the segment being used to access the offsets and functions appropriate for the specific segment.    TOAST's recursion requires no such tracking, since the algorithm depends solely upon the corners of the pixels being subdivided.

In practice, it is a fairly trivial matter to output a particular TOAST tile using GPU vertex and index buffers in such 3D systems as Direct3d, OpenGL and WebGL. For efficiency, the tile pyramid is always evaluated from the root tile. This is a single tile comprising of 8 triangles.  For each triangle vertex there are coordinates in 3 space (on the sphere), and in the 2D image space. As we noted above, for each tile, the corners of the tile footprint and the directionality of the triangle bisecting the quad are determined from parent tile, and the tile triangle is recursively subdivided a number of times. Each triangle is subdivided into four triangles with the new points being calculated as the geometric mean of the edge segments.

The tile has four quadrants representing the coverage of that tile, and four vertex and index buffers. Each tile can draw either its entire contents, or any combination of its quadrants, or have one of its children draw itself in place of that quadrant. In three dimensions, these coordinates are treated as vectors and normalized so they are on the surface of a unit sphere. In two dimensons, image coordinates are the 2D midpoint of the segments in image space where the units correspond to the UV texture coordinates for the image tile in the range of 0-1.

Once the recursion is done to create a mesh of sufficient density there is minimal error in final projection of the texture image for that tile. Tiles are evaluated to ensure two conditions: 1) are they visible in the view frustum and 2) are they of sufficient size to represent image textures at approximately 1:1 ratio. Tiles that are not in the view frustum are culled. Tiles that are not of sufficient image density have their children render that quadrant of the image. Having the root tile draw itself, will have the entire visible image drawn and projected onto the screen with the proper resolution.

Unless the entire image pyramid is already loaded, there is an intermediate step where a parent tile knows it needs its child tile to draw, but the image data is not yet available. The parent tile requests a tile cache management service to query a tile and draw the currently available lower resolution data until the correct resolution data can be downloaded and initialized by the tile cache manager.

The tile cache manager can handle background downloading, prioritizing the queue, and removing old tiles from memory when they have not been rendered for a long time.

TOAST is not the only projection that meets the requirement for densely rendering the sky in a single rectangle.  We have encountered several others in this paper and there are doubtless more.  TOAST was used in the WWT for a combination of historical and algorithmic advantages.  The availability of the SDSS catalog data in HTM and the clear existing definition of the recursive process were attractive at the time of WWT's creation.  In practice the recursive nature of the determination of pixel boundaries where the next generation is defined not in terms of some global function but in terms of the current set of pixel locations is very convenient.  Having great circle boundaries for the pixels is also very helpful when assessing the edges of each pixel.

% * <tom.mcglynn@nasa.gov> 2018-04-18T21:15:43.154Z:
%
% ^.
\section{Discussion}
Figures \ref{fig:TOT}, \ref{fig:TEA} and \ref{fig:TOA} show images of the sky with a coordinate grid in three realizations of the octahedral projections.  In all cases there are significant features at the boundaries of the original tiles as our experience with the cubic projections led us to expect.  The TOAST projection is more rounded towards the poles than the either the TOT or TEA projections. In each there are discontinuities in the slopes of lines even at the pole due to the squashing of the equilateral octahedron facets.

 The unconventional placement of the pole at the center of the projection, and the features at the boundaries may make these projections less suitable for direct display than more conventional all sky projection.  This has little bearing on their utility as intermediate representations.

The TTN and TEA projections are mathematically straightforward, have continuous derivatives except at the tile boundaries and are easily invertible.  The TOAST projection is much more difficult to define in general, but it is particularly straightforward to calculate when a grid of $2^n \times 2^n$ pixels (or some contiguous subset thereof) is to be computed.  For this special case -- which happens to correspond precisely to the needs for hierarchical imaging -- the TOAST projection is very easily computed.  Essentially only a single square root function needs to be evaluated for each point.

The WorldWide Telescope uses TOAST as its projection for storing survey data.  The \textit{SkyView} Virtual Telescope\footnote{http://skyview.gsfc.nasa.gov} supports all three of the projections defined here.  

The FITS WCS conventions \citep{Calabretta2002} use a three-letter abbreviation for each projection.  We have suggested TOT for the Triangular Octahedral Tangent plane projection, TEA for the Triangular octahedral Equal Area projection, and TOA for the Tessellated Octahedral Adaptive spherical transformation (TOAST) projection.  The modified Cahill projection shown in Figure \ref{fig:CAH-proj} has also been implemented in \textit{SkyView}.  By analogy with the TSC projection we suggest the abbreviation TSO for Tangential Spherical Octahedron projection.

The utility of the cube and octahedron may suggest that we might wish to consider the other regular solids for projections. 
% * <tom.mcglynn@nasa.gov> 2018-05-04T14:35:51.412Z:
% 
% Modified to add reference to Tegmark 1996.
% 
% ^.
\citet{Tegmark1996} has contemplated using the icosahedron as a basis for a partitioning into hexagonal pixels.  However we have seen no obvious natural approach that would enable us to transform this or the other geometric solids into a projection that meets our original criteria.

\acknowledgements

We would like to thank Drs Aniruddha Thakar and Alex Szalay for their help in providing figures for HTM triangulation of the sphere and pointers to the Geomview package which remains a very useful tool despite the limited support it has had for many years.  Conversations with Dr Gregory McGlynn were very helpful in understanding the differentiability of the TOAST projection.  The \textit{SkyView} system to illustrate the projections in the paper has been supported by a series of NASA Astrophysics Data Program and Astrophysics Applied Information Systems Research Program grants and is now hosted at NASA's High Energy Astrophysics Science Archive Research Center (HEASARC).  We gratefully acknowledge NASA's support.
% * <tom.mcglynn@nasa.gov> 2018-04-27T20:14:24.911Z:
% 
% Moved WWT stuff to appendix
% 
% ^.
\appendix
{\bf  WorldWide Telescope Conventions}

By convention in WorldWide Telescope and most other TOAST systems, TOAST image pyramids consist of multiple levels of tiles. Each tile is a 256x256 image in equatorial coordinates for sky images and panoramas, the North celestial pole is in the center, RA 0h on the right, 6h on top, 12h on the left, and 18h on the bottom. For Planetary surfaces the 0 degrees is left, 90 degrees on the bottom, 180 degrees on the right and 270 degrees at the top.

Tiles are quad-trees and are conventionally accessed by either a triple of level, x, and y, or a quad-tree key.  The first tile level is zero and consists of the root tile, with a coordinate of 0, 0. At each successive level the tile count doubles on each axis. A tile key is empty for level 0, then for each new level of subdivision, the address of the quadrant ID for each level deeper in the subdivision is appended to the key. For example, a tile with address level=2, x=3, y= 3, the quad-tree key would be 33. 

Tile trees are assumed to be accessible on the Internet through a URL. The URL access pattern can be provided to a TOAST browser application through a pattern string that allows the substitution of the tile address for either level, x, and y, the tile key, or by pattern substitution found in the pattern substitution table. Then the image at the calculated URL can be downloaded. If the image does not exist, then that means there is no further data available for that quadrant or below it in the pyramid.

\begin{lstlisting}[caption=Example 1: WTML file][H]
<?xml version='1.0' encoding='UTF-8'?>
  <Folder Name="WWT" Group="View">
    <ImageSet 
Generic="False" 
DataSetType="Sky" 
BandPass="Visible" 
Name="Digitized Sky Survey (Color)"     
      Url="http://cdn.worldwidetelescope.org/wwtweb/dss.aspx?q={L},{X},{Y}" 
BaseTileLevel="0" 
TileLevels="12" 
BaseDegreesPerTile="180" 
FileType=".png" 
BottomsUp="False" 
      Projection="Toast" 
QuadTreeMap="0123" 
CenterX="0" CenterY="0" 
      OffsetX="0" OffsetY="0" 
Rotation="0" 
Sparse="False" 
ElevationModel="False" 
StockSet="True">
      <ThumbnailUrl>http://www.worldwidetelescope.org/thumbnails/DSS.png
</ThumbnailUrl>
<Credits>Copyright DSS Consortium</Credits>
    <CreditsUrl>http://gsss.stsci.edu/Acknowledgements/DataCopyrights.htm</CreditsUrl>
   </ImageSet>
</Folder>
\end{lstlisting}

WorldWide Telescope uses a XML format called WTML\footnote{A full definition of WTML is available at http://www.worldwidetelescope.org/docs/WorldWideTelescopeDataFilesReference.html.} that defines a dataset for TOAST (see Example 1), once a TOAST pyramid is calculated and made available on the Internet. A description of the metadata can be communicated to a TOAST browser through a WTML ImageSet definition.

In this example XML, a 12-level deep (approximately 1 Terapixel image), of .png file type is defined.  Substitution parameters (enclosed in braces) can be placed anywhere in the URL to rewrite the URL to refer to the tile needed for access. This allows both statically mapped tiles in a filesystem, or programmatically accessed tiles to be used as sources.
\bibliography{toast.bib}
\end{document}